\documentclass[twocolumn,amsmath,aps,superscriptaddress,showpacs,floatfix,a4paper]{revtex4}
\usepackage{amsmath} 
\usepackage{graphicx} 
\usepackage{epsf}
\usepackage{bm}  
\usepackage{latexsym} 
\RequirePackage{amsopn}
\RequirePackage{amssymb} 
\RequirePackage{amsfonts}
\RequirePackage{amsthm} 
\usepackage[]{graphicx} 
\usepackage[]{amsmath}

\newcommand{\lb }{{\langle}}
\newcommand{\rb}{{\rangle}}     

\begin{document} 

\title{Hydrodynamics of confined colloidal fluids in two dimensions }
\author{Jimaan San\'e}
\affiliation{Rudolf Peierls Centre for Theoretical Physics,
           1 Keble Road, Oxford OX1 3NP, United Kingdom}
\affiliation{Department of Chemistry, Cambridge University, 
           Lensfield Road, Cambridge CB2 1EW, United Kingdom}
\author{Johan T.\ Padding}
\affiliation{Computational Biophysics, University of Twente, 
           PO Box 217, 7500 AE, Enschede, The Netherlands}
\author{Ard A.\ Louis}
\affiliation{Rudolf Peierls Centre for Theoretical Physics,
           1 Keble Road, Oxford OX1 3NP, United Kingdom}
\date{\today}

\begin{abstract}
  We apply a hybrid Molecular Dynamics and mesoscopic simulation
  technique to study the dynamics of two dimensional colloidal discs
  in confined geometries.  We calculate the velocity autocorrelation
  functions, and observe the predicted $t^{-1}$ long time hydrodynamic
  tail that characterizes unconfined fluids, as well as more complex
  oscillating behavior and negative tails for strongly confined
  geometries.  Because the $t^{-1}$ tail of the
  velocity autocorrelation function is cut off for longer times in
  finite systems,  the
  related diffusion coefficient does not diverge, but instead depends
  logarithmically on the overall size of the system.
\end{abstract}

\maketitle

\section{Introduction}

The role of hydrodynamics in two dimensions (2d) is considerably more
complex than in three dimensions (3d).  For example, when, in 1851,
George Gabriel Stokes tried to extend his famous calculation of the
low Reynolds (Re) number flow field around a sphere~\cite{Stokes} to
that of a cylinder he found that~\cite{Lamb32}

\begin{quote}
The pressure of the cylinder on the fluid continually tends to increase
the quantity of fluid which it carries with it, while the friction of the
fluid at a distance from the cylinder continually tends to diminish it. In
the case of a sphere, these two causes eventually counteract each other, and
the motion becomes uniform. But in the case of a cylinder, the increase in the 
quantity of fluid carried continually gains on the decrease due to the friction
of the surrounding fluid, and the quantity carried increases indefinitely as the
cylinder moves on.
\end{quote} 
so that there was no finite solution.  This was later called the ``Stokes Paradox''.
Experimental realizations of 2d systems are, of course, always
embedded in one way or another in the 3d world.  In a classic paper,
Saffman~\cite{Saff76} demonstrated  how taking into account the upper and lower
boundaries on a 2d system solves the Stokes Paradox because these
boundaries open up a new channel for momentum flow out of the system.
If the viscosity of the confining medium is $\eta'$, while the
viscosity of the confined medium of height $h$ is $\eta$, then a new
length scale emerges:
\begin{equation}\label{SaffmanL}
L_S \sim \frac{h \eta}{\eta'},
\end{equation}
beyond which the true 3d nature of the whole system needs to be taken into
account.  The zero Re number Stokes equations also cease to be valid at
distances larger than  $L_{\mbox{Re}} \sim \nu/U$, where $\nu$ is the kinematic
viscosity and $U$ the velocity of the fluid,  because inertial forces must be
taken into account. Although inertial terms also  become relevant at similar
length-scales in 3d, this fact doesn't need to be taken into account to obtain
bounded solutions of the  Stokes equations.  For length scales $L \lesssim
min\{L_S,L_{\mbox{Re}}\} $ the total momentum in the 2d layer is approximately
conserved and Saffman showed that for a disk of radius $R_c$ and thickness $h$,
the 2d diffusion coefficient for stick boundary conditions takes the following
finite form~\cite{Saff76}: 
\begin{equation}\label{SaffmanD}
D^{2d} = \frac{k_B T}{4 \pi \eta h} \left[ \ln \left(\frac{h \eta}{R_c
    \eta'} \right) - \gamma \right].
\end{equation}
where $k_B$ is Boltzmann's constant, $T$ is the temperature, and
$\gamma = 0.5572$ is Euler's constant.  Note that in contrast to the
3d form, where the diffusion coefficient only depends on $k_B T$,
$R_c$ and $\eta$, here both the thickness of the film $h$ and the
viscosity of the boundary $\eta'$ enter into the expression for the
diffusion coefficient. Eq.~(\ref{SaffmanL}) also implies that 2d
hydrodynamics will be most evident when the confining boundary has a
very low viscosity.

Examples of experimental
systems where 2d hydrodynamics are important include diffusion of protein and
lipid molecules in biological
membranes~\cite{Cone72,Cone73,Cone74,Edidin74,Pete82}.  Cicuta \emph{et
al}~\cite{Cicu07} recently directly measured the diffusion of liquid
domains in giant unilamellar vesicles (GUVs) and found that the mean square
displacement of the domains scaled logarithmically with their radius, in
agreement with Saffmans prediction. 

Experiments on colloidal particles confined in a thin sheet of fluid (such as a
soap film) have used video imaging~\cite{Cheu96} and optical
tweezers~\cite{DiLe08} to explicitly demonstrate 
that the hydrodynamic interaction between the particles decays
logarithmically with distance.  These effects can be understood from
solving the 2d Stokes equations and carefully taking into account the
boundary conditions.  Because the 3d boundary in these cases is air,
with a much smaller viscosity than the soap solution, $L_S$ can be as
large as $0.1m$ or more.  The low Re numbers typical of colloidal
suspensions mean that $L_{Re}$ can be much larger than that, on the
order of many meters.  Furthermore, if the 2d systems under
investigation has boundaries at a distance $L \ll
min\{L_{\mbox{S}},L_{\mbox{Re}}\}$ then the diffusion coefficient
scales with system size as~\cite{Happ73}
\begin{equation}\label{eqD2d}
D \sim \ln\left[L/R_c\right].
\end{equation}

The goal of this paper is to use computer simulations to study the
hydrodynamics of colloidal discs in confined geometries.  We limit
ourselves to two dimensions (2d), which has the advantage that
simulations in 2d are faster than in 3d.  The price we pay for this is
that we must take into account some of the subtleties of 2d
hydrodynamics described earlier, such as the finite size effects
illustrated, for example, by Eq.~(\ref{eqD2d}).  But  these
effects can also be observed in experiments on quasi-two dimensional
systems, and are therefore interesting in their own right.

We use a combination of Stochastic Rotation Dynamics (SRD)
~\cite{Male99,Ihle03,Padd06} to describe the solvent, and Molecular
Dynamics to solve the equations of motion for the colloids. Such a
hybrid technique was first employed by Malevanets and Kapral
~\cite{Male00}, and used to study colloidal sedimentation by
ourselves~\cite{Padd04} and by Hecht \textit{et al.}~\cite{Hech05}.
We have recently completed an extensive study of this method to study
the hydrodynamics of colloidal suspensions~\cite{Padd06}, which we
will call ref I, and  we summarize some of the main points of the
method in Section~\ref{sec_method}.

Particle based methods like SRD (note that in the literature this
method is also sometimes called Multiple Particle Collision Dynamics,
see e.g.~~\cite{Ripo05}). have the advantage that boundary conditions
are very easy to implement as external fields.  This contrasts with
traditional methods of computational fluid dynamics  where
boundary conditions are typically harder to implement.  This suggest
that methods like SRD may be ideally suited for the study of colloids
in confined geometries. The rapid development of new methods to create
microfludic systems is also stimulating experimental studies on
colloids in confined geometries~\cite{Squi05}. For
that reason, computer simulation techniques that can calculate the
properties of colloids in narrow channels will become increasingly
important.  Another field of possible application includes flow in
porous media~~\cite{Edo03,Edo05}.

We proceed as follows:  In section~\ref{sec_method} we describe the
hybrid Molecular Dynamics/SRD method we employ,  and sketch out the key
hydrodynamic parameters that govern the flow behavior.   Section III
describes simulations of a pure SRD fluid system in 2d, where we find
that the effects of hydrodynamic correlations are more pronounced than
those found in 3d~\cite{Ripo05}. We also explore the important role of
finite size effects.  In Section IV we calculate the velocity
autocorrelation function for colloids in 2d, and show how confinement
qualitatively affects their long-time behavior.  In Section V, we analyze the
diffusion coefficient for colloids in 2d, and show how the confinement
effects seen for the velocity auto-correlation function are connected
to the behavior of the diffusion coefficient. We summarize our main conclusions in section VI.

\begin{figure*}
\begin{center} 
\includegraphics[angle=-90, width=0.75\textwidth]{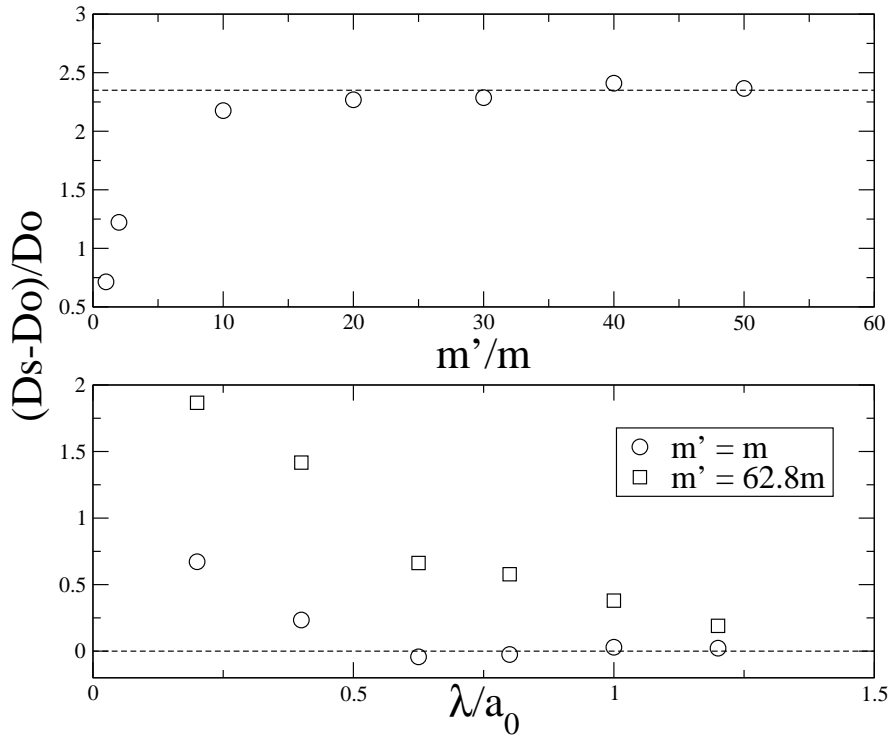}
\caption{\label{Fig:weightsrd} Top : Deviation of the simulated diffusion
coefficient $D_{s}$, from the random collision approximation $D_o$ predicted
by Eq.~(\ref{eq:tracerdiff}), as a function of the heavy particle mass. We simulated
fluid particles in 2d for a square geometry with walls separated by
a distance 
$L=32a_{0}$. 
Bottom : Deviation of the simulated diffusion coefficient $D_{s}$, from the
random collision approximation $D_o$, as a function of the particle mean free
path $\lambda$.  Simulations were performed for values of
$\lambda = 0.2,0.4,0.625,0.8,1,1.2$.}
\end{center}
\end{figure*}

\section{Hybrid MD-SRD coarse-grained simulation method}\label{sec_method}

To describe the hydrodynamic behavior of colloids, induced by a
background fluid of much smaller cons\textbf{t}ituents, some form of
coarse-graining is required. The hydrodynamics can be described by the
Navier Stokes equations that coarse-grain the fluid within a
continuum description.  The downside of going directly through this
route is that every time the colloids move, the boundary conditions on
the differential equations change, making them computationally
expensive to solve.  

An alternative to direct solution of the Navier Stokes equations is to
use particle based techniques that exploit the fact that only a few
conditions, such as (local) energy and momentum conservation, need to
be satisfied to allow the correct (thermo) hydrodynamics to emerge in
the continuum limit. Simple particle collision rules, easily amenable
to efficient computer simulation, can therefore be used.  Boundary
conditions (such as those imposed by colloids in suspension) are
easily implemented as external fields.  One of the first methods to
exploit these ideas was direct simulation Monte Carlo (DSMC) method of
Bird~~\cite{Bird70,Bird94}.  The Lattice Boltzmann (LB) technique
where a linearized and pre-averaged Boltzmann equation is discretized
and solved on a lattice~~\cite{Succ01}, is a popular modern
implementation of these ideas, and in particular has been extended by
Ladd and others to model colloidal
suspensions~~\cite{Ladd93,Ladd01,Cate04,Loba04,Capu04,Chat05}.

In this paper we implement the SRD method first derived by Malevanets
and Kapral~~\cite{Male99}. It resembles the Lowe-Anderson
thermostat~\cite{Lowe99}, but has the advantage that transport
coefficients have been analytically
calculated~\cite{Ihle03,Kiku03,Pool04}, greatly facilitating its use.
It is important to remember that for all these particle based methods,
the  particles should not be viewed as some kind of composite
supramolecular fluid units , but rather as coarse-grained Navier Stokes
solvers (with noise in the case of SRD)~\cite{Padd06}.

An SRD fluid is modeled by $N$ point particles of mass $m$, with
positions ${\bf{r}}_{i}$ and velocities ${\bf{v}}_{i}$. The coarse
graining procedure consists of two steps, streaming and collision. During
the streaming step, the positions of the fluid particles are updated via
\begin{equation}
{\bf{r}}_{i}(t+\delta t_{c}) = {\bf{r}}_{i}(t) + {\bf{v}}_{i}(t)\delta t_{c}.
\end{equation} In the collision step, the particles are split up into
cells with sides of length $a_0$, and their velocities are rotated
around an angle $\alpha$ with respect to the cell center
of mass velocity,
\begin{equation}
{\bf{v}}_{i}(t+\delta t_{c}) = {\bf{v}}_{c.m,i}(t) +
\mathcal{R}_i(\alpha)\left[{\bf{v}}_{i}(t)-{\bf{v}}_{c.m,i}(t)\right]
\end{equation}
where ${\bf{v}}_{c.m,i}=\sum^{i,t}_{j}(m{\bf{v}}_{j})/\sum_{j}m$ is the center
of mass velocity of the particles the cell to which $i$ belongs,
$\mathcal{R}_i(\alpha)$ is the cell rotational matrix and $\delta t_{c}$ is the
interval between collisions. The purpose of this collision step is to transfer
momentum between the fluid particles while conserving the energy and momentum
of each cell.

The fluid particles only interact with one another through the
collision procedure. Direct interactions between the solvent particles
are not taken into account, so that the algorithm scales as
$\cal{O}(N)$ with particle number. This is the main cause of the
efficiency of simulations using SRD. The carefully constructed
rotation procedure can be be viewed as a coarse-graining of particle
collisions over space {\em and} time.  Mass, energy and momentum are
conserved locally, so that on large enough length-scales the correct
Navier Stokes hydrodynamics emerges, as was shown explicitly by Malevanets $\&$
Kapral ~\cite{Male99}.

An advantage of SRD is that it can easily be coupled to a solute as first shown
by Malevanets and Kapral~\cite{Male00}, and studied in detail in a recent
paper by two of the present authors~\cite{Padd06} (ref I).  If we wish to simulate the behavior of spherical
colloids of mass $M$, they can be embbeded in a solvent using a Molecular
Dynamics technique. For the colloid-colloid interaction we use a
standard steeply
repulsive potential of the form:
\begin{displaymath}
\varphi_{cc}(r) = \left\{
\begin{array}{ll}
4\epsilon\left( \left(\frac{\sigma_{cc}}{r}\right)^{48}-
\left(\frac{\sigma_{cc}}{r}\right)^{12} +\frac{1}{4}\right) & (r\leq 2^{1/24}\sigma_{cc})\\
0 & (r\geq 2^{1/24}\sigma_{cc})
\end{array} \right.
\end{displaymath}
while the interaction between the colloid and the solvent is described by
a similar, but less steep, potential:
\begin{displaymath}
\varphi_{cs}(r) = \left\{
\begin{array}{ll}
4\epsilon\left( \left(\frac{\sigma_{cs}}{r}\right)^{12}-
\left(\frac{\sigma_{cs}}{r}\right)^{6} +\frac{1}{4}\right) & (r\leq 2^{1/6}\sigma_{cs})\\
0 & (r\geq 2^{1/6}\sigma_{cs})
\end{array} \right.
\end{displaymath}
where $\sigma_{cc}$ and $\sigma_{cs}$ are the colloid-colloid and
colloid-solvent collision diameters.  We propagate the ensuing equations of
motion with a Velocity Verlet algorithm~\cite{AllenTildesley} using a molecular
dynamic time step $\Delta t$
\begin{eqnarray} R_{i}(t+\Delta t) &=&
  R_{i}(t) + V_{i}(t)\Delta t + \frac{F_{i}(t)}{2M}\Delta t^{2}\\
  V_{i}(t+\Delta t) &=& V_{i}(t) + \frac{F_{i}(t)+F(t+\Delta
    t)}{2M}\Delta t 
\end{eqnarray} 
where $R_{i}$ and $V_{i}$ are the position and velocity of the colloid, and
$F_{i}$ the total force exerted on the colloid.  Coupling the colloids in this
way leads to slip boundary conditions.  Stick boundary conditions can also be
implemented~\cite{Padd05}, but for qualitative behavior, we don't expect there
to be important differences.  In parallel the velocities and positions of the
SRD particles are streamed in the external potential given by the colloids and
the external walls and updated with the SRD rotation-collision step every
time-step $\delta t_c$.

To prevent spurious depletion forces, we set the interaction range
$\sigma_{cf}$ slightly below half the colloid diameter $\sigma_{cc}/2$
and include a small compensating potential for very short distances
(when $\beta \varphi_{cc}(r) \geq 2.5$). For further details of how
this procedure reproduces the correct equilibrium behavior see Ref
I~\cite{Padd06}.

The larger the ratio $\sigma_{cc}/a_0$, the more accurately the hydrodynamic
flow fields will be reproduced.  Here we use $\sigma_{cc}/a_0 = 4.3$, and
$\sigma_{cf} = 2 a_0$, which was shown in ref I to reproduce the flow fields
with small relative errors for a single sphere in a 3d flow.  Other parameters
choices taken from Ref I include $\epsilon_{cc}=\epsilon_{cf} = 2.5 k_BT$ for
the colloids, and $\gamma =5, \alpha = \frac12 \pi$ for the SRD particle number
density and rotation angle respectively.  The time-steps for the MD and SRD
step are set by slightly different physics~\cite{Padd06}, and we chose $\Delta
t = 0.025 t_0$ and $\delta t_c = 0.1 t_0$, where $t_0 = a_0 \sqrt{\frac{m}{k_B
T}}$ is the unit of time in our simulations.

Coarse-graining methods like SRD are useful when they make the
calculation of certain desired physical properties more efficient.  To
achieve this, compromises must be made (there is no such thing as a
free lunch).  For colloidal suspensions, for example, the Re number is
typically very low, on the order of $10^{-5}$ or less, and similarly
the Mach number Ma $= U/c_s$, where $U$ is a typical system velocity
and $c_s$ is the velocity of sound, can be as small as $10^{-10}$.  To
achieve this in a particle based simulation is extremely expensive.
Resolving sound waves would mean that, since they travel much faster
than colloidal particles, extremely small time-steps would be
necessary in the simulation.  Luckily even for Ma numbers as high
$0.1$ the hydrodynamics can be accurately approximated by
incompressible hydrodynamics, so that one doesn't need to fulfill the
physical condition to obtain essentially the same physics. Similarly,
for many applications, as long as the Re number is significantly lower
than 1, the system can still be accurately described by the Stokes
equations.  A more detailed discussion of these length-scales and
hydrodynamic numbers can be found in ref I, and we will implicitly be
making use of these arguments for the current work.

A similar set of arguments can be made for the time-scales of a real
colloidal fluid, compared to those found in our coarse-grained
description.  For example the kinematic time, defined as $\tau_\nu =
\sigma_{cs}^2/\nu$, i.e.\ the time it takes a the vorticity to diffuse one
colloidal radius, is of order $10^{-6}$s for a buoyant colloid of
radius $1 \mu m$ suspended in water.  For the same system, the
diffusion time $\tau_D = \sigma_{cc}^2/D \approx 5s$.  Resolving these
time-scales in one simulation would be very inefficient.  In ref I we
claim that successful coarse-graining techniques must telescope down
the hierarchy of time-scales to a more manageable separations that are
efficient for computational purposes.  We argue that what is needed is
not an exact representation of all the time-scales of the physical
system, but rather clear time-scale separation.  For example, having
$\tau_\nu$ be only one or two orders of magnitude smaller than
$\tau_D$ can still lead to an accurate description of the desired
physics.  However, interpreting the results means taking this
telescoping down of time-scales into account, and to do this properly,
one has keep careful track of the physics involved.  Expressing
results as much as possible in terms of dimensionless units can
facilitate this process~\cite{Padd06}.

\section{Dynamics of solvent particles}

\begin{figure*}
\begin{center}
\includegraphics[angle=-90, width=0.75\textwidth]{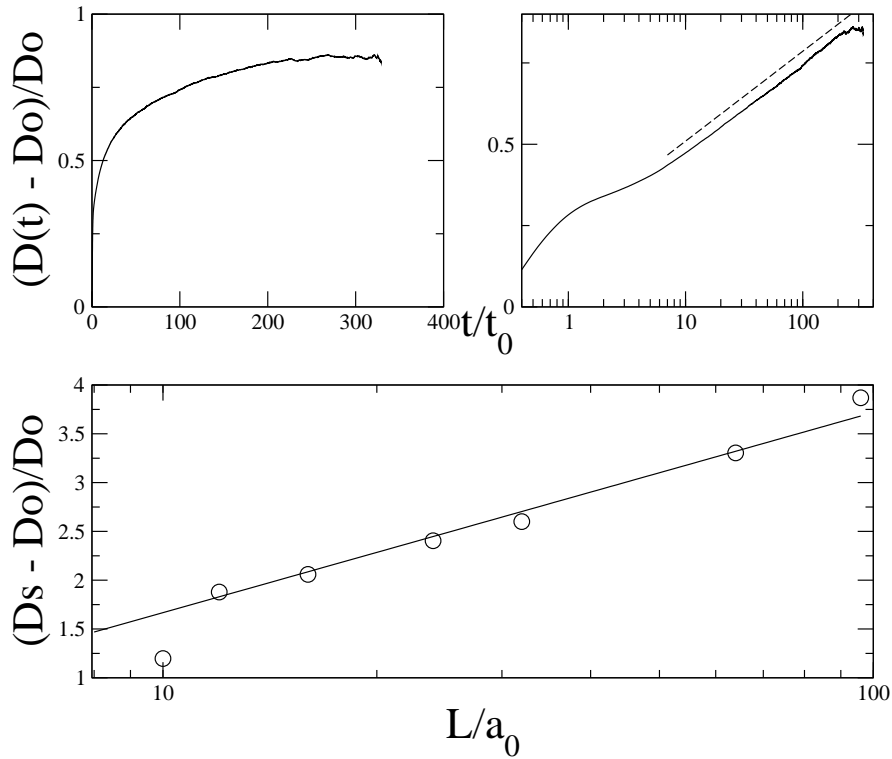}
\caption{\label{Fig:diffsrd} The top two plots show the temporal evolution of
the self diffusion coefficient of a fluid particle in a large box of size $256 a_0 \times 256 a_0$ with periodic boundary conditions. The right plot shows that rather
than saturate, the diffusion coefficient grows as $D \sim \ln t$, as expected
from theory.  The bottom plot shows the hydrodynamic corrections to the
diffusion coefficient $D$ compared to to the random collision approximation
expression $D_0$ given by Eq~(\protect\ref{eq:tracerdiff}) for different box
sizes $L$. As expected, these corrections shows a logarithmic growth with
$L/a_o$.}
\end{center}
\end{figure*}

Before investigating the behavior of colloids in suspension, we study
a simpler problem of an SRD fluid confined to two dimensions.   Much
of this section will follow on an earlier comprehensive study by Ripoll et
al.~\cite{Ripo05} in 3d, but here we focus on 2d. 

We begin by deriving an expression for the velocity autocorrelation
function of the SRD particles, following similar steps to those found
in ref.~\cite{Ripo05} for 3d.
The $n^{th}$ collision step of the SRD method can be rewritten as
\begin{eqnarray}\nonumber\label{eq:col}
{\bf{v}}_{i}(n\delta t_{c}) &=& {\bf{v}}_{i}((n-1)\delta t_{c}) +
(\mathcal{R}_i(\alpha)-\mathcal{I})\\
&\times&\left[{\bf{v}}_{i}((n-1)\delta t_{c})-{\bf{v}}_{c.m,i}((n-1)\delta t_{c})\right]
\end{eqnarray}
where $\mathcal{I}$ is the unit matrix, and $t=n\delta t_{c}$ the discretized
time, with $n$ the number of collision steps, $\delta t_{c}$ the collision
interval and ${\bf{v}}_{c.m,i}$ the cell center of mass velocity. The rotation
matrix is defined in two dimensions as
\begin{displaymath}
\mathcal{R}_i(\alpha)  = \left( \begin{array}{cc}
\cos\alpha & \pm \sin\alpha \\
\mp\sin\alpha &\cos\alpha
\end{array} \right)
\end{displaymath}
such that the rotational average over any vector ${\bf{A}}$ becomes
\begin{equation}\label{eq:rot}
\lb (\mathcal{R}(\alpha)-\mathcal{I}){\bf{A}}\rb  =
-(1-\cos\alpha){\bf{A}}
= -\zeta_{\alpha}\lb {\bf{A}}\rb .
\end{equation}
If we now assume density fluctuations in each cell to be small,
we can write $\lb {\bf{v}}_{c.m,i}(n\delta t_{c})\rb \simeq
\frac{1}{m\gamma}\lb {\sum_{j}^{i,n}\bf{v}}_{j}\rb$. By multiplying each
side by $\lb{\bf{v}}_{i}(0)\rb$ and further assuming the velocity
of colliding particles to be uncorrelated, we arrive
at
\begin{equation}\label{eq:molchaos}
\lb {\bf{v}}_{c.m,i}((n-1)\delta t_{c}){\bf{v}}_{i}(0)\rb \simeq
\frac{1}{m\gamma}\lb {\bf{v}}_{i}((n-1)\delta t_{c}){\bf{v}}_{i}(0)\rb.
\end{equation}
where $\gamma$ is the average number of solvent particles per cell.
Substituting (\ref{eq:molchaos}) and (\ref{eq:rot}) into (\ref{eq:col}), and
rearranging, we obtain an expression for the correlation of a fluid particle
\begin{equation}\label{eq:col2}
\lb {\bf{v}}_{i}(n\delta t_{c}){\bf{v}}_{i}(0)\rb
=(1-\zeta_{\alpha}\zeta_{\rho}^{m})
\lb {\bf{v}}_{i}((n-1)\delta t_{c}){\bf{v}}_{i}(0)\rb
\end{equation}
where $\zeta_{\rho}^{m} = 1-1/\gamma$. This expression shows that we can write
the correlation at a certain time step in terms of the previous time step,
such that the normalized velocity autocorrelation function (VACF) is,
\begin{equation}
\frac{\lb {\bf{v}}_{i}(n\delta t_{c}){\bf{v}}_{i}(0)\rb }{\lb
{\bf{v}}^{2}_{i}(0)\rb }\simeq\zeta^{n}
\end{equation}
where $\zeta=1-\zeta_{\alpha}\zeta_{\rho}^{m}$ is the decorrelation
factor. The VACF, for reasons that will become apparent later, is
the quantity of interest here and has the form
\begin{equation}\label{eq:decor}
\lb {\bf{v}}_{i}(n\delta t_{c}){\bf{v}}_{i}(0)\rb \simeq \frac{k_{B}T}{m}\zeta^{n}
\end{equation}

A similar analysis can be performed for the case of a single heavy
tracer particle of mass $m'$ embedded in a solvent~\cite{Ripo05}. The
total mass in a collision box is then $(M+m\gamma)$ such that the
center of mass correlation is written as
\begin{equation}\label{eq:Mmolchaos}
\lb {\bf{v}}_{c.m,i}(n\delta t_{c}){\bf{v}}_{i}(0)\rb \simeq\frac{m'}{m\gamma+M}
\lb {\bf{v}}_{i}(n\delta t_{c}){\bf{v}}_{i}(0)\rb .
\end{equation}
By substituting (\ref{eq:Mmolchaos}) into (\ref{eq:col2}),  the
decorrelation factor for a heavy tracer particle is found to be
\begin{equation}\label{eq:zeta}
\zeta = 1-\zeta_{\alpha}\frac{m\gamma}{m\gamma+m'}=1-\zeta_{\alpha}\zeta_{\rho}^{M}.
\end{equation}

The self diffusion constant $D$ of a particle $i$ is related to its
mean square displacement via the Einstein relation~\cite{Hansen86}:
\begin{equation}\label{DD}
D = \lim_{t\to\infty}\frac{1}{4
t}\lb[{\bf{r}}_{i}(t)-{\bf{r}}_{i}(0)]^{2}\rb.
\end{equation}
The position of a particle can be written explicitly in terms of
discrete time-steps \begin{equation} 
{\bf{r}}_{i}(t) = {\bf{r}}_{i}(0) + \delta t_{c}\sum^{n-1}_{k=0}{\bf{v}}_{i}(k\delta t_{c}),
\end{equation}
so that
\begin{equation} \lb[{\bf{r}}_{i}(t)-{\bf{r}}_{i}(0)]^{2}\rb=\delta t_{c}^{2}\sum^{n-1}_{j=0}\sum^{n-1}_{k=0}
\lb {\bf{v}}_{i}(j\delta t_{c}){\bf{v}}_{i}(k\delta t_{c})\rb.
\end{equation}
We note that combining the equation above with Eq.~({\ref{DD}}) leads
to the discrete form of the standard Green-Kubo expression for the
diffusion coefficient as an integral over the velocity autocorrelation function.
Manipulating the sums, we find~\cite{Tuzel03}
\begin{eqnarray}\label{eq:sums}
\sum^{n-1}_{j=0}\sum^{n-1}_{k=0}
&\lb& {\bf{v}}_{i}(j\delta t_{c}){\bf{v}}_{i}(k\delta t_{c})\rb\\\nonumber
&=& \sum^{n-1}_{j=0} \lb {\bf{v}}_{i}^{2}(j\delta t_{c})\rb + 
2 \sum^{n-2}_{j=0}\sum^{n-1}_{k=j+1}
\lb {\bf{v}}_{i}(j\delta t_{c}){\bf{v}}_{i}(k\delta t_{c})\rb\\\nonumber
&=& 2n\frac{k_{B}T}{m} + 2\sum^{n-1}_{j=1}j
\lb {\bf{v}}_{i}(0){\bf{v}}_{i}((n-j)\delta t_{c})\rb .
\end{eqnarray}
Substituting the expression for the VACF derived earlier (\ref{eq:decor}) into
(\ref{eq:sums}), we can write the diffusion coefficient in terms of its
decorrelation factor $\zeta$,
\begin{equation}\label{eq:diffco}
D = \lim_{n\to\infty}\frac{k_{B}T}{m}\delta t_{c}\left[\frac{1}{2}
+ \frac{1}{n}\sum^{n-1}_{j=1}j\zeta^{n-j}\right]
= \frac{k_{B}T\delta t_{c}}{2m}\left[\frac{1+\zeta}{1-\zeta}\right].
\end{equation}
Substituting Eq.\ref{eq:zeta} into Eq.\ref{eq:diffco}, results in the
following dimensionless expressions for the self diffusion constant of a fluid
and  heavy tracer particle respectively 
\begin{eqnarray}\label{eq:tracerdiff}
\frac{D_{0}^{m}}{D_0} &=&
\lambda\left[\frac{1}{1-\cos\alpha}\left(\frac{m\gamma}{m\gamma-1}\right)
-\frac{1}{2}\right]\\
\frac{D_{0}^{m'}}{D_0} &=& \frac{\lambda
m}{m'}\left[\frac{1}{1-\cos\alpha}\left(\frac{\gamma+\frac{m'}{m}}{\gamma}\right)
-\frac{1}{2}\right].
\end{eqnarray}
$D_0$ denotes the unit of diffusion and is expressed as
$a^2_0/t_0=a_0\sqrt{k_BT/m}$ and $\lambda$ is the dimensionless mean free path.
It is a measure of the average distance the fluid particles  travel in between
collisions and has the form~\cite{Padd06} 
\begin{equation}
\lambda = \frac{\delta t_{c}}{a_{0}}\sqrt{\frac{k_{B}T}{m}} = \frac{\delta t_{c}}{t_{0}}.
\end{equation}

These expressions for $D$  make a key approximation, namely that collisions are always
random, and that the particle velocities are uncorrelated.  This neglects any
hydrodynamic effects.  These expressions are thus expected to become more
accurate if the mean-free path $\lambda$ becomes larger so that the random
collision approximation is expected to be a better description.  Ripoll
\emph{et al} ~\cite{Ripo05} showed that in 3d, for their simulation parameters,
the expression~(\ref{eq:tracerdiff}) for self-diffusion of an SRD
particle began to show
significant deviations from measured values when the mean-free path was smaller
than $0.6$.  Similarly, they found that for smaller mean-free paths $\lambda
=0.1$, these expressions could underestimate the diffusion coefficient
of a tagged heavier particle of mass $M$ by as
much as $75\%$ for $M \geq 10m$.

In Fig.~\ref{Fig:weightsrd} we analyze the self-diffusion coefficient
of a tagged SRD particle as a function of mass and of mean free path
for a square geometry with plates $L=32 a_0$ SRD cell widths
wide. Similarly to Ripoll {\em et al.}~\cite{Ripo05} we find
deviations due to hydrodynamics, but in 2d these are much more
pronounced.  For example, as the mass increases, the hydrodynamic
corrections to Eq.\ref{eq:tracerdiff} saturate at a deviation of
over $200 \%$ for larger masses.  Similarly, we observe larger
deviations as a function of mean free path than found in 3d.

In contrast to the 3d results, for which finite size effects are not
very strong, we expect that in 2d the effect of box size will be much
more pronounced.  To illustrate this, we carried out simulations in a
much larger square box of width $L=256a_{0}$ box sizes, now with periodic
boundary conditions.  These are shown in the top two plots of
Fig.~\ref{Fig:diffsrd}.  We observe that the temporal diffusion
coefficient, defined as
\begin{equation}\label{eq:diffconst-t}
D(t) = \int^{t}_{0}\lb v(t')v(0)\rb dt',
\end{equation}
continues to grow with time in a manner consistent with the expected
scaling $D \sim \ln[t]$, as illustrated in the   second top plot in
Fig.~\ref{Fig:diffsrd}.  We expect the diffusion coefficient to
eventually saturate for this finite box size.  But for  an infinite
box, we expect that  $D(t)$
will continue to grow indefinitely, a manifestation of the Stokes Paradox.
Similarly, for a  fixed box  of size $L^2$, we expect that $D \sim
L/a_0$~\cite{Happ73}, as discussed in the introduction, and this scaling is
indeed observed in the bottom panel in  Fig.~\ref{Fig:diffsrd}.

\section{Velocity autocorrelation functions of colloidal particles}

\begin{figure*}
\begin{center}
\includegraphics[angle=-90, width=0.75\textwidth]{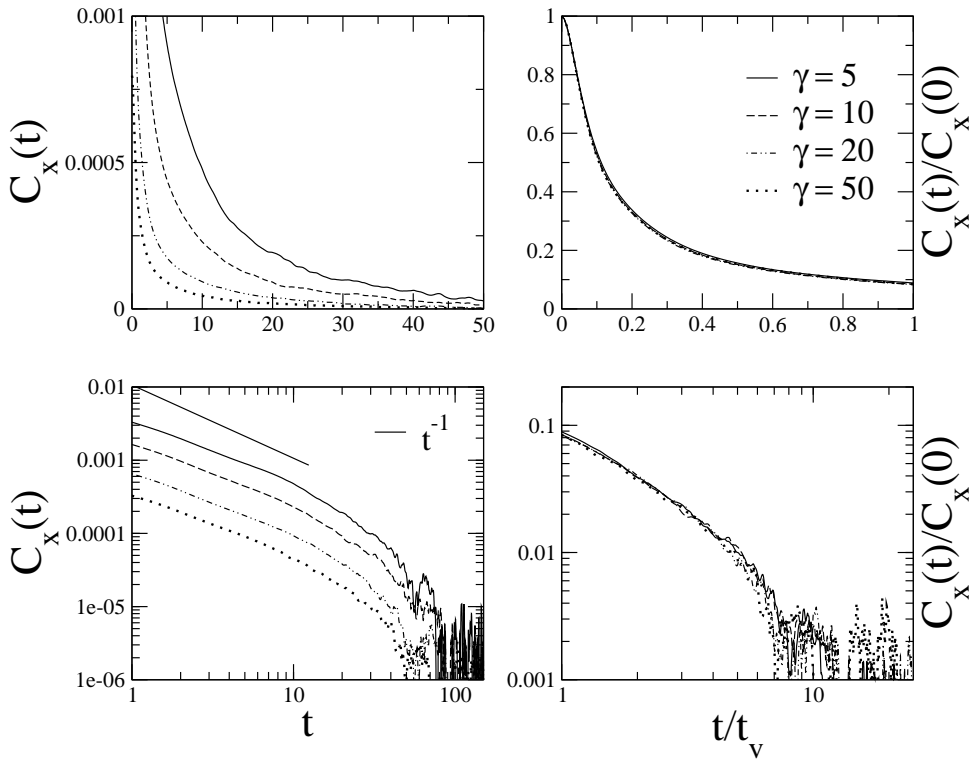}
\caption{\label{Fig:vacttailsgamma}
Scaling of the velocity autocorrelation function : when VACF is normalized and
plotted in terms of the reduced time $t/t_{\nu}$, all the data collapse to the
same curve. The VACF was originally measured for varying solvent
densities $\gamma$. The
system size is  $L=32a_{0}$, which implies that $\chi_s = L/2\sigma_{cs} = 8$.}
\end{center}
\end{figure*}

\begin{figure}
\begin{center}
\includegraphics[angle=-90, width=0.5\textwidth]{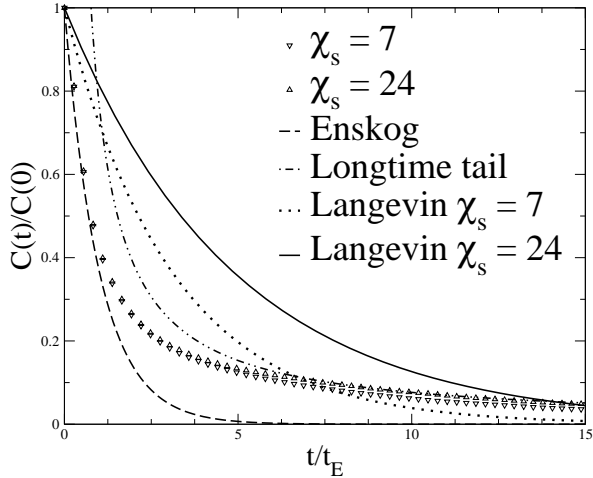}
\caption{\label{Fig:Enskog1} Normalized colloid VACFs simulated for increasing
pipe width with $\chi_s = \frac{L}{2\sigma_{cs}}=7,24$, with $L=28,96a_0$.
Simulations results were scaled with the Enskog time is $t_{E}=1.0888t_{0}$
calculated from (\ref{eq:Enskogth}). The dashed line represents the short time
decay from (\ref{eq:Enskog}), the dashed-dotted line represent the decay from
the long time tail calculated from (\ref{eq:kinetic}) whilst the dotted line
and the solid line denote the Langevin decay with the friction $\xi$ for the
two respective box sizes.}
\end{center}
\end{figure}

\begin{figure}
\begin{center}
\includegraphics[angle=-90,width=0.5\textwidth]{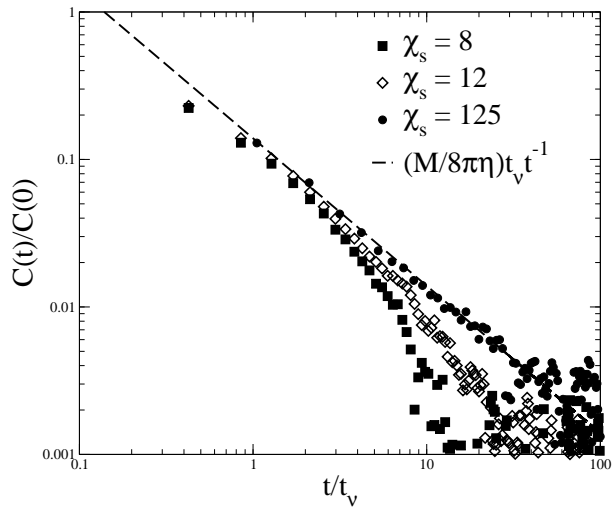}
\caption{\label{Fig:tailstw} Log-log plot of the VACF for of a colloidal
particle in boxes of size $L=32,48,500a_0$ ($\chi_{s}=8,12,125$).
The dashed line is the fit from Eq.\ref{eq:kinetic} and serves as a guide
to the eye.}
\end{center}
\end{figure}

\begin{figure*}
\begin{center}
\includegraphics[angle=-90,width=0.75\textwidth]{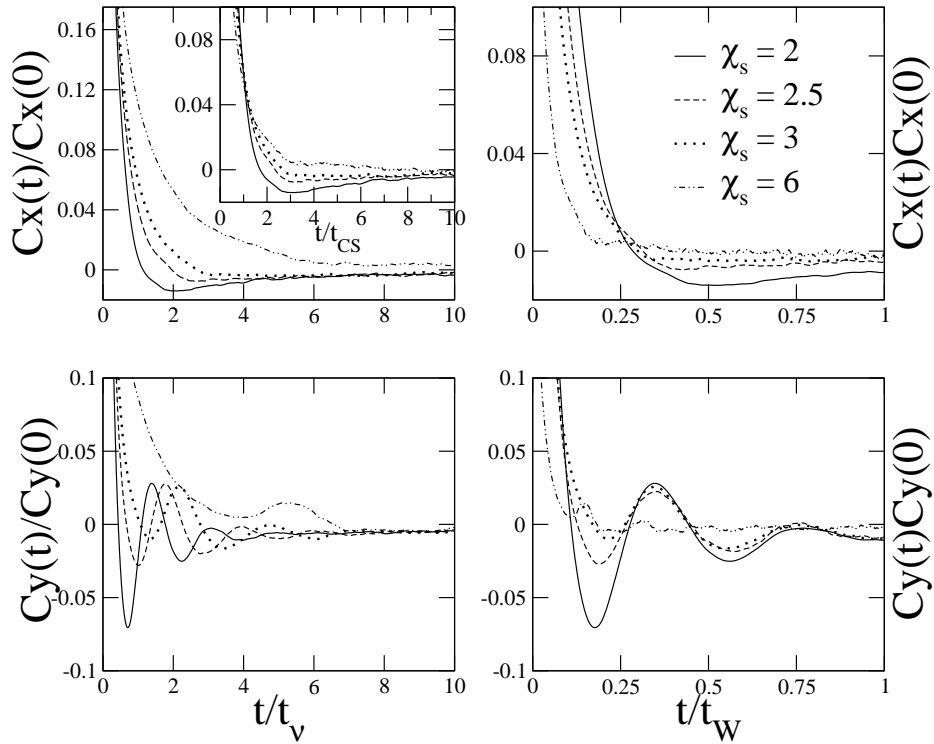}
\caption{\label{Fig:normtails1} Normalized velocity autocorrelation
function of a colloidal particle confined between two walls for increasing
values of $\chi_s= L/2\sigma_{cs}$.  Simulations here were performed for pipe
widths $L=8,10,12,24a_0$. The different plots denote the components of the
normalized VACF parallel $C_x(t)$ (top) and perpendicular $C_y(t)$ (bottom) to
the walls.  Note that all plots are scaled with the kinematic time $t_{\nu} =
\sigma_{cs}^{2}/\nu$, and the time-scale for momentum to diffuse the distance
between the walls $t_{w} = L^{2}/4\nu$.  The minimum of the negative tails
observed for $C_x(t)$ scale on top of each other when time is scaled with the
sonic time $t/t_{cs}$ (insert in upper right panel).  Similarly, the
oscillating tails for $C_y(t)$ show the same period when scaled with $t/t_W$
(bottom right panel).} 
\end{center}
\end{figure*}

\begin{figure*}
\begin{center}
\includegraphics[angle=-90,width=0.75\textwidth]{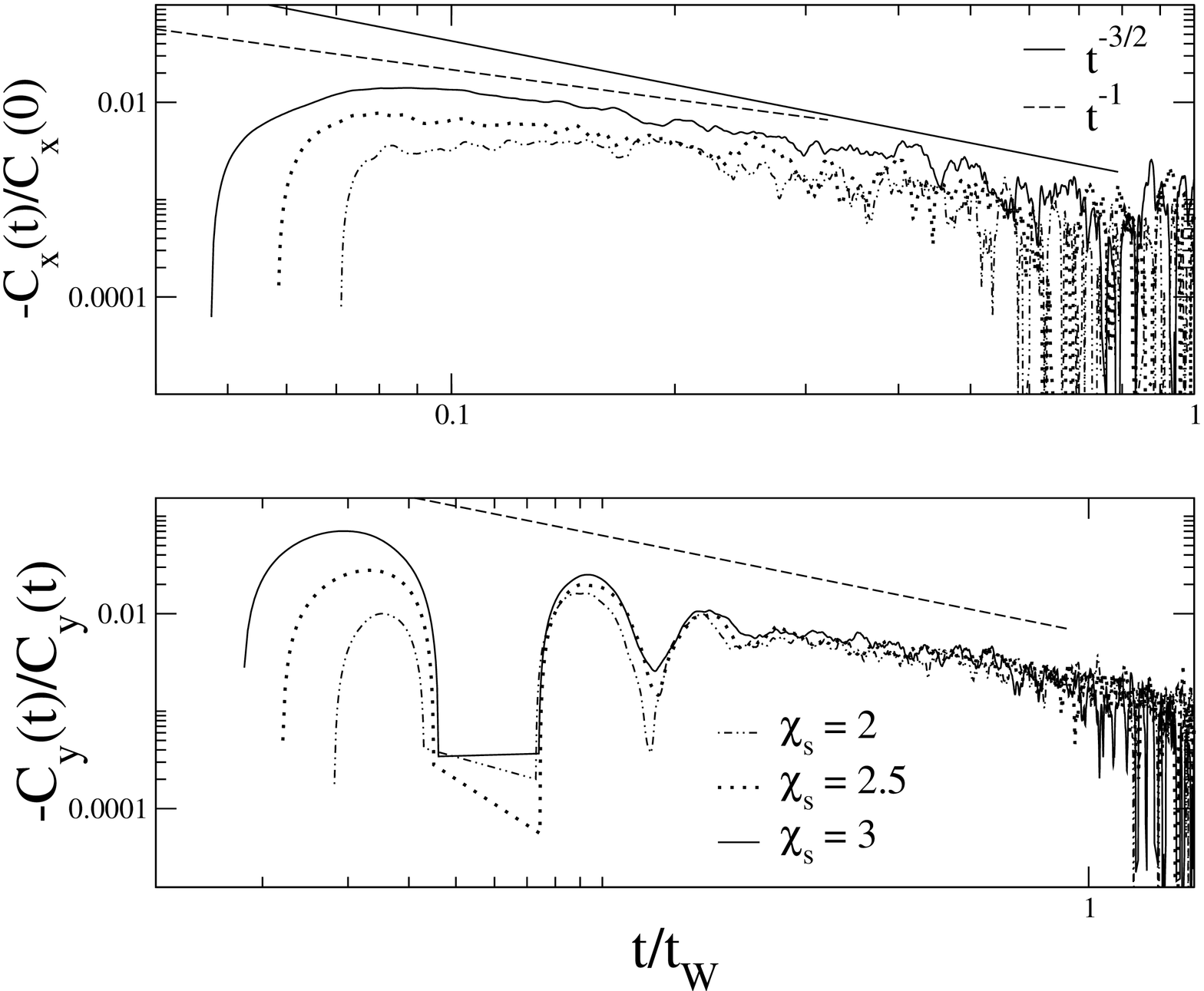}
\caption{\label{Fig:normtails2} Log-log plot of the VACF of a
  confined colloidal
particle for boxes of size $L=8,10,12a_0$ (
$\chi_{s}=2,2.5,3$ respectively).
The dashed line and the straight line have slopes $t^{-1}$ and
$t^{-3/2}$ respectively. Only the positive values of $-C(t)$ 
are plotted here. }
\end{center}
\end{figure*}

Having worked out some properties of diffusing SRD particles, we now
turn to the properties of colloidal particles embedded in a solvent.

If memory effects are ignored in a simple Langevin equation description of a
spherical colloid of mass $M$, then the  velocity autocorrelation function
(VACF) of a colloidal particle can be calculated to be~§~\cite{Russ89}
\begin{equation}\label{eq:expdec}
\lb v(t)v(0)\rb  = \frac{k_{B}T}{M}\exp(-t/t_{\xi}),
\end{equation}
where the time $t_{\xi} = M/\xi$ indicates how quickly
particles forget their initial velocity. Its integral is related to the
diffusion coefficient through the Einstein relation:
\begin{equation}\label{eq:Einstein}
D = \int^{\infty}_{0}\lb v(t)v(0)\rb dt 
= \frac{k_{B}T}{\xi}.
\end{equation}
The Einstein relation is of course valid for any physical description of
the VACF.

Langevin approaches have traditionally been used for colloidal systems when
hydrodynamics could be ignored. 
However, it is well known that hydrodynamic
effects can have an important qualitative effect on the VACF.  In their
pioneering work, Alder and Wainwright~\cite{Alde70} used MD simulations to
demonstrate that the VACF ($C(t)$)  of a tagged particle exhibits an algebraic
decay at long times of the form $t^{-d/2}$, instead of the exponential form
predicted by the Langevin equation.   They showed that this behavior was a
consequence of momentum conservation, and therefore quite general.  For
colloidal particles in 3d, the diffusion coefficient is dominated by the
contributions from this long time tail~\cite{Padd06}, and we expect the same to
be true in 2d.  The correlation function for a colloid with slip boundary
conditions can be calculated from kinetic theory ~\cite{Alde70}:
\begin{equation}\label{eq:kinetic}
\langle v(t)v(0)\rangle  =
\left(\frac{d-1}{d\rho}\right)\frac{k_{B}T}{(4\pi(D+\nu)t)^{d/2}},
\end{equation}
where $d$ is the number of dimensions, and $\rho$ the solvent density.
This calculation predicts a $t^{-1}$ power for the tail in 2 dimensions.  That this
should cause problems for the definition of $D$ is evident from
Eq.\ref{eq:Einstein} because it implies that D diverges
logarithmically with time.  Note that similar behavior was seen for
pure SRD particles in Fig.~\ref{Fig:diffsrd}, where we found the scaling $D(t) \sim
\ln[t]$.  For the colloids, we expect that the tail in the VACF will
form on the timescale $t_{\nu}=\sigma_{cs}^{2}/\nu$ it takes the
kinematic viscosity $\nu$ to diffuse over the particle radius.

Fig.\ref{Fig:vacttailsgamma}  shows simulations run for a  square box with a width $L
= 32a_0$.  Eq.\ref{eq:kinetic} predicts that the tail
should scale as $(( \nu +D) \rho t)^{-1}$.  We tested this further by varying
the number density $\gamma$ and simultaneously changing the density of the
colloids so that they remain buoyant.  For SRD the kinematic viscosity $\nu$
depends only very weakly on $\gamma$~\cite{Pool04,Padd06} for large values of
$\gamma$ and keeping in mind that from equipartition 
\begin{equation}\label{eq:V(0)}
C(0) = \lb v(0)^2\rb = \frac{k_B T}{M}
\end{equation}
it is not hard to show that the long time tails should all scale onto the same
curve if time is scaled with $t/t_\nu$. We show this
explicitly in Fig.~\ref{Fig:vacttailsgamma} for a fixed system size.  

At times shorter than the kinematic time, there is a contribution to the
overall diffusion that comes from the local random collisions between the
colloid and the solvent particles. This is typically dominant on time scales
less than the sonic time $t_{cs} = v_s/\sigma_{cc}$ over which collective modes
can be generated~\cite{Padd06}.  We can calculate it using standard
Enskog kinetic theory, and the ensuing Enskog friction coefficient $\xi_{E}$ has
the following form~\cite{Phd}
\begin{equation}\label{eq:Enskogth}
\xi_{E}^{2d} = \frac{3\sqrt{2}}{4}\sigma_{cs}\gamma\pi^{3/2}\left(
k_BT\frac{mM}{m+M}\right)^{1/2}
\end{equation}
in two dimensions. Thus for very short times, the decay of the
VACF is characterized by the Enskog time
$t_{E}=M/\xi_{E}^{2d}$ and it follows that
\begin{equation}\label{eq:Enskog}
\langle v(t)v(0) \rangle = \frac{k_BT}{M}\exp(-t/t_{E}).
\end{equation}
because the collisions are essentially random.

As shown in Fig.~\ref{Fig:Enskog1}, for 
 short times, on the order of the Enskog time $t_{E}$, the autocorrelation
function shows clear exponential decay, in good agreement with
Eq.\ref{eq:Enskog}.   The simulations shown are for two box sizes,
and for short times, the VACF are independent of system size, as
expected from Enskog theory.   

At longer times, Fig.~\ref{Fig:Enskog1} clearly shows the beginning of
the long-time tail.  The theoretical line we plot is from
Eq.~\ref{eq:kinetic}, and fits remarkably well to the data.
However, we note that there are some small deviations with system size
at these longer times, which will be explained below.

We also note that a direct comparison with the Langevin equation shows
that for short times the Langevin equation overestimates the VACF, and
that for longer times in underestimates the VACF for colloids.  A more
in depth discussion of this point can be found in appendix B of Ref I.
In addition, in two dimensions, the Langevin
equation~(\ref{eq:expdec}) would predict an exponential form with
different $t_\xi$ for different box sizes, because the diffusion
coefficient changes with box size.  By contrast, our results show that
for short times the VACF is independent of box size.
Clearly the Langevin equation does a poor job in capturing details of
the colloidal VACF.

\begin{figure*}
\begin{center}
\includegraphics[angle=-90, width=0.75\textwidth]{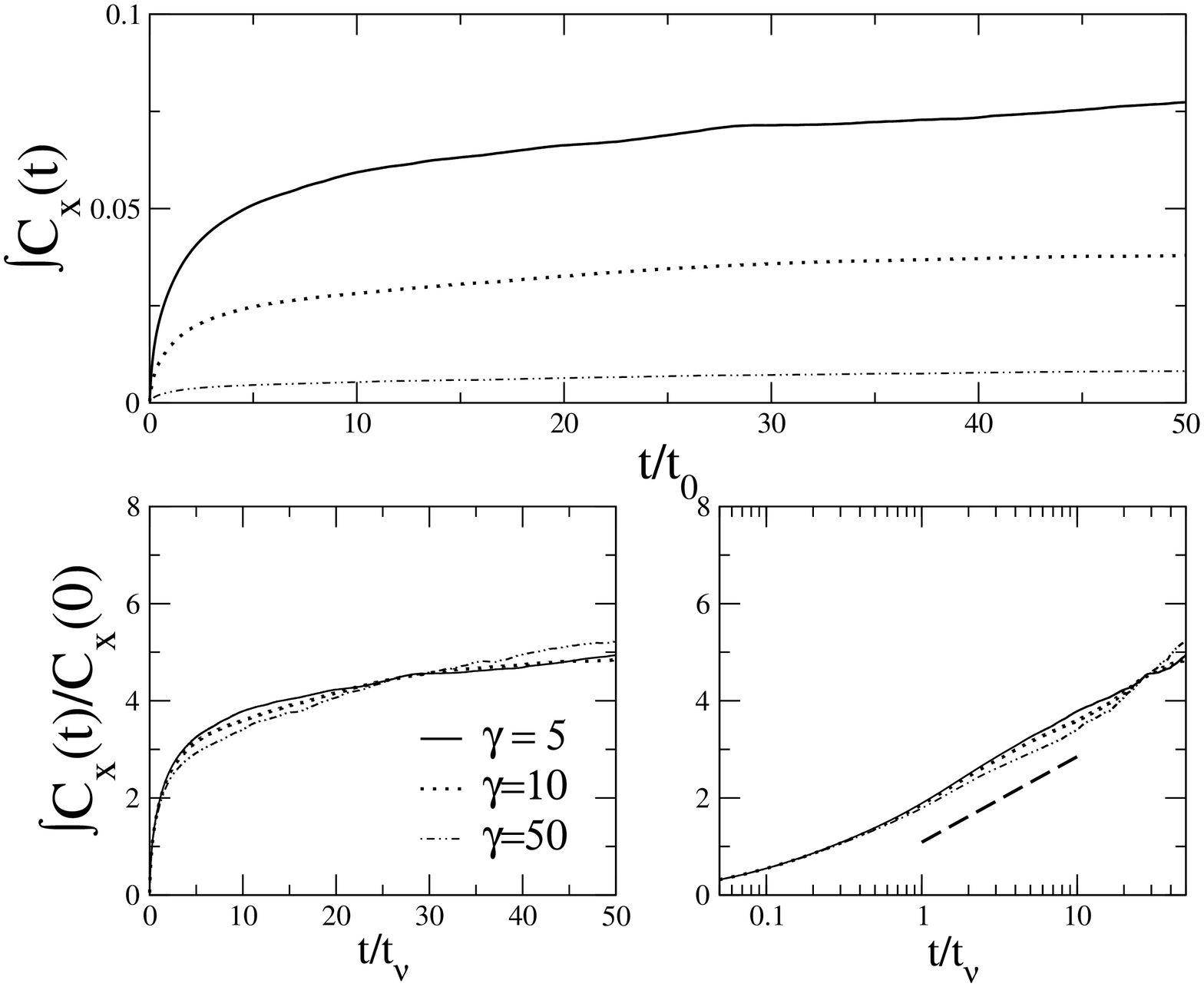}
\caption{\label{Fig:fitgamatnu2} Time evolution of the time-dependent diffusion
  coefficient  of a buoyant colloid in two dimensions for SRD particle
densities $\gamma = 5, 10, 50$.  The bottom two graphs show the diffusion
coefficient normalized by $C_x(0) = k_B T/M$, with time scaled by $t/t_\nu$.
We expect the graphs to scale on to each other for longer times where the
long-time tails dominate. (see also Fig.~\protect\ref{Fig:vacttailsgamma}).  The
bottom right graph  has a logarithmic scale and the dashed line has the slope
$\frac{M}{8\pi\eta}$ and serves as a guide to the eye. }
\end{center}
\end{figure*}

While the simulations above are for fixed boundaries, it is also interesting to
see what happens to the VACF when the confinement is more pronounced.  In
confined geometries, the particle induced flow fields should feel the presence
of the walls.  Bocquet and Barrat~\cite{Boc} showed that a sink in the decay of
the long time tails should occur after an observation time on the order of
$t_{w}=\frac{L^{2}}{4\nu} = \chi_s^2 t_\nu$. This time is characteristic of 
how long it takes for the kinematic viscosity to reach the wall, when $L/2$ is the
average distance to the wall.   We illustrate the effect of the wall on the
VACF in Fig.~\ref{Fig:tailstw} for three different box sizes.   For the two
narrower boxes, the VACF clearly begins to drop below the $t^{-1}$ power law
but for the largest box, of size $L=500a_0$, we don't observe any deviation within
our error bars.  The sink in the tail for the $\chi_s=8$ simulation run begins at
an observation times less than $10t_{\nu}$, whereas  the kinematic wall time in
this instance is $t_W\approx 64 t_{\nu}$. That may be because of other
wall effects that kick in earlier for such a narrow box, or it may be that the
cutoff in the algebraic decay is gradual and commences sooner than predicted by
Bocquet and Barrat.

In an important study, Hagen \emph{et al} ~\cite{Pago01} used Lattice
Boltzmann simulations to investigate the VACF of a colloidal particle
between rigid walls and found qualitative deviations from the standard
long-time tails.  In particular, for a sphere in a narrow enough cylinder, they
found negative tails for the VACF $C_x(t)$ parallel to the walls
that exhibited an algebraic decay like $C_x(t) \sim t^{-3/2}$.
Similarly, for a two dimensional disc between two plates they found
$C_x(t) \sim t^{-3/2}$, and for a three dimensional sphere between two
plates they found $C_x(t) \sim t^{-2}$.  These exponents depend on the
confinement, rather than on the overall dimension of the system. They
explained the emergence of this negative tail with a simple
mode-coupling theory that takes into account the fact that the sound
wave generated by the colloid becomes diffusive. They further noticed
that for slip walls, the normal behavior was recovered, suggesting
the origin of the negative tail lies in the existence of velocity
gradients near the wall.

We performed simulations of colloidal discs in a pipe of length
$512a_{0}$ with periodic boundaries in the $x$ direction and with two
stick boundary condition walls at a reduced distance $\chi_{s} = L/2\sigma_{cs}
= 2,2.5,3,6$, apart in the $y$ direction and show the results in Fig.~\ref{Fig:normtails1}.  We find
a negative tail for $C_x(t)$, the VACF parallel to the plates, We find that the
amplitude of the negative tail grows with increasing confinement.  Furthermore,
when time is scaled with $t/t_{cs}$, the different correlation functions all
show a minimum at about $t \approx 3 t_{cs}$, suggesting that sound waves are
indeed the dominant cause of the negative tail, as suggested in ~\cite{Pago01}.
For the larger confinement shown here, $\chi_{s} = 6$, the VACF does show a
rapid decay, but there does not seem to be a negative tail.  This
suggest that the diffusive sound wave mechanisms are still playing a part
in the smallest ($\chi_{s}=8$) simulations of Fig.~\ref{Fig:tailstw},
and may explain why the VACF decays on a shorter time $t/t_W$ than
predicted by Bocquet and Barrat \cite{Boc}.

In the bottom two panels of Fig.~\ref{Fig:normtails1} we observe
oscillatory behavior for $C_y(t)$, the VACF perpendicular to the
plates.  This can be explained as follows: when a particle moves in
the $y$ direction towards the wall, it sets up a momentum flow which
can reflect off the wall and come back a time later to push the
particle in the opposite direction.  This effect should become more
pronounced for stronger confinement, as we observe.  To check this
mechanism, we note that the walls introduce another length-scale, $t_W
= \frac{L^2}{4\nu}$, which is the time it takes vorticity to diffuse to the
walls. If this reflection mechanism is at play, we would expect the
period of the oscillations to reflect this time-scale.  In the bottom
right panel of Fig.~\ref{Fig:normtails1}, we observe that when $C_y(t)$
is scaled with the time $t/t_W$ the oscillation minima indeed fall on
top of each other, at least for sufficiently strong confinement.

 As discussed by Hagen \emph{et al}~\cite{Pago01}, the
$C_x(t)$ should exhibit a negative tail that scales like $t^{\frac{-3}{2}}$ for
sufficiently strong confinement.  In  the upper plot of
Fig.~\ref{Fig:normtails2}  we indeed observe that the exponent is
greater than $t^{-1}$,  and consistent with $t^{-\frac32}$, as expected,
although our data is not clean enough to confirm the exact exponent.
Similarly, the final decay of the component $C_y(t)$ appears closer to
$t^{-1}$ than to $t^{-\frac32}$.

Clearly confinement has an important effect on the long-time behavior
of the VACF, and there may be further subtle effects that we have not
yet been uncovered.  It would be interesting, for example, to see how
the angular correlation functions, studied in ref.~\cite{Padd05} with
SRD for 3d stick boundary colloids in the bulk phase, would behave
under confinement. However, for the calculation of long-time tails,
methods like Lattice Boltzmann techniques used by Hagen \emph{et
  al.}~\cite{Pago01}, where noise does not play a big role, may be
simpler and faster to use.

\section{Diffusion coefficient of colloidal particles under confinement}

\begin{figure*} \begin{center} \includegraphics[angle=-90,
width=0.75\textwidth]{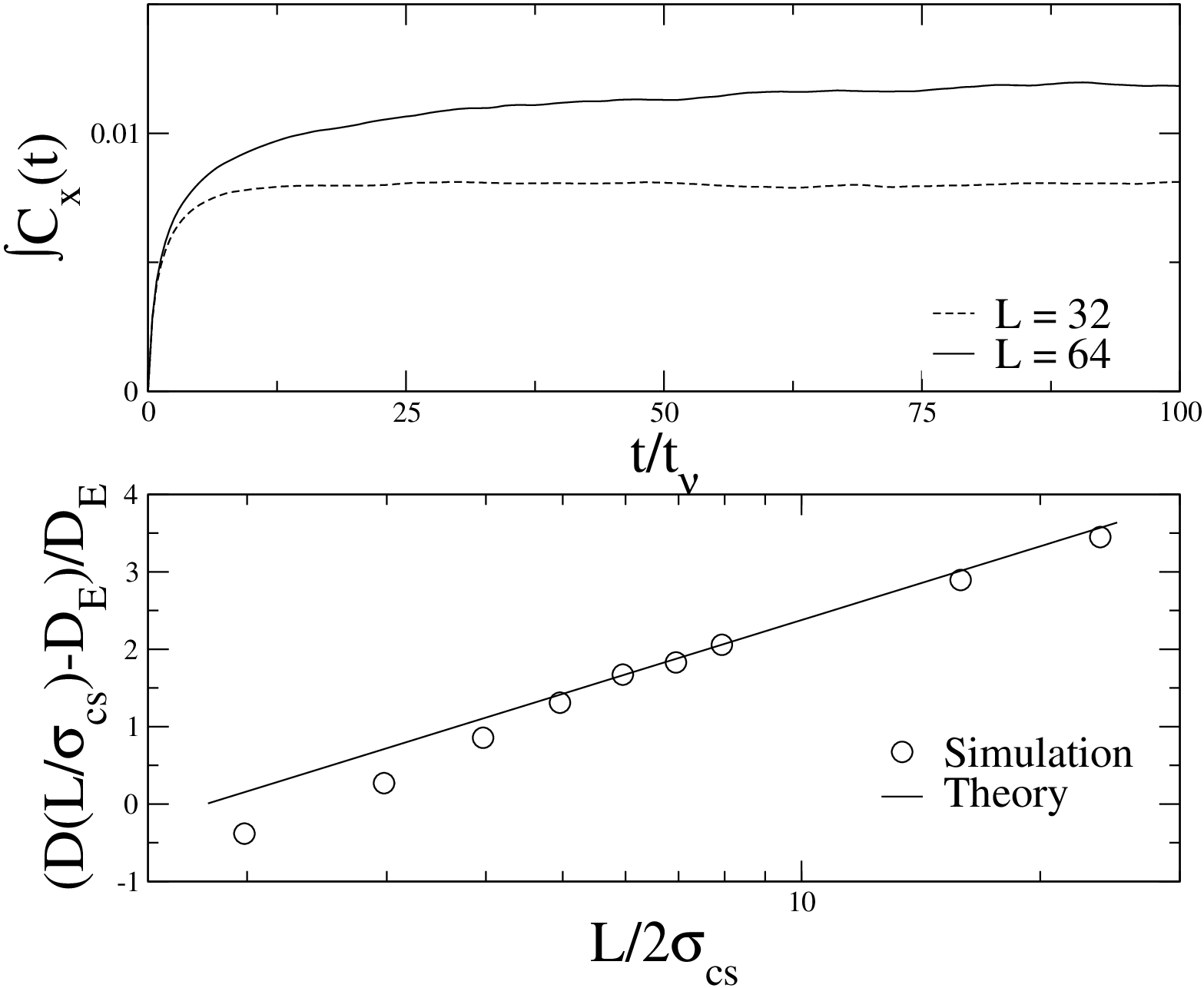} \caption{\label{Fig:diffpipe222} Top : The
effect of confinement on diffusion.  The two curves represent the temporal
evolution of the self diffusion coefficient for colloids in varying
confinement. The dashed line denotes colloids diffusing between plates a
distance $L=32a_{0}$ apart ($\chi_s=8$) whilst in the case of the solid line
the plate separation is $64a_{0}$ ($\chi_s=16$).  For the smaller pipe, the
kinematic wall time is $t_W\approx 64 t_{\nu}$.  The diffusion coefficient
begins to plateau much earlier than that because the effect of the wall on the
VACF kicks in earlier than that.  Note that only the x-component of the
diffusion is plotted here.  Bottom : Simulations were performed for 2d pipes of
respective widths $L=8,12,16,20,24,28,32,64,96a_{0}$. The straight line is the
fit from (\ref{eq:pipe}) and the circles are the simulation points.  }
\end{center}
\end{figure*}

The Einstein relation~(\ref{eq:Einstein}) directly relates the VACF
and the diffusion coefficient.  We found that for short times, the
VACF was well described by an Enskog form~(\ref{eq:Enskog}) that was
largely independent of the boundaries, and that at longer times it
exhibited a long time tail that was much more sensitive to the
boundaries.  For strong confinement, the tail could even be negative
or oscillatory, but for weak confinement, it appears to scale as $C(t)
\sim t^{-1}$.

For an unbounded 2d system, the diffusion coefficient does not converge,
instead its behavior with time can be approximated as:
\begin{flushleft}
\begin{eqnarray}\nonumber
D^{2d}(t) &=& \int^{t}_{0}\lb v(t)v(0)\rb dt\\\nonumber
&\approx& \frac{k_{B}T}{M}\left(\int^{t_{\nu}}_{0}\exp(-t/t_{E})dt\right)
+ \int^{t}_{t_{\nu}}\frac{k_{B}T}{8\pi\rho\nu t}  \\\label{eq:timediff}
&\approx& \frac{k_BT}{\xi^{2d}_{E}} + \frac{k_{B}T}{8\pi\eta}
\left[\ln t\right]^{t}_{t_{\nu}} 
\end{eqnarray}
\end{flushleft}
where we have assumed that the Enskog and hydrodynamic contributions
to the VACF can be separated (this is not quite true) and moreover
that the hydrodynamic tail does not kick until a time scale on the
order of the kinematic time $t_\nu$. We also assume that $D \ll \nu$.

\subsection{Simulations in the 'bulk'}

In Fig.~\ref{Fig:fitgamatnu2}, we present the temporal evolution of the self
diffusion coefficient of a colloid for a large box. We approximated colloids in
the bulk by using a box of size $L^{2} = 256a_0 \times 256a_0$ with periodic
boundary conditions.  The plot shows results for solvent densities
$\gamma=5,10,50$.  On the time-scales of the simulation, we observe behavior
consistent with $D \sim \ln[t]$,  as expected from the $t^{-1}$ tail of the
VACF.  In practice this would mean that $D$ would grow indefinitely with time,
and be unbounded, which is a manifestation of the Stokes Paradox.

\subsection{Simulations in confinement}

Whereas the diffusion coefficient of a two dimensional disc in the bulk
appears to grow in an unbounded fashion with time, the diffusion
coefficient for a confined fluid is expected to saturate at a finite
value~\cite{Happ73,Boc}.  We showed in
Figs~\ref{Fig:vacttailsgamma}~-~\ref{Fig:normtails2} that the VACF
 is affected by the presence of walls, and no longer shows the $t^{-1}$
 behavior at very long times that would lead to a logarithmic
 divergence. As a result of the wall interaction, the diffusion will no longer
diverge, but will plateau at a value determined by the distance to the wall.

We tested this simple argument by simulating colloids under two different
levels of confinement. The top panel of Fig.~\ref{Fig:diffpipe222} shows the
integral of the velocity autocorrelation function plotted for colloids
diffusing between parallel plates a distance $L=32a_{0}$ and $L=64a_{0}$ apart
respectively. For the smaller system, the temporal diffusion
coefficient reaches a plateau at shorter times than is found for the
larger system.

To make these arguments more quantitative, we make the following
approximation to the diffusion coefficient:
\begin{eqnarray}\label{eq:pipe}\nonumber
D^{2d}(L/\sigma_{cs}) &\sim& \int^{t_{W}}_{0}\lb v(t)v(0)\rb dt \\\nonumber
&\approx& \frac{k_{B}T}{\xi_{E}^{2d}} + \frac{ k_{B}T}{8\pi\eta}
\left[\ln t\right]^{t_{W}}_{t_{\nu}} \\\nonumber
&=& D_E + \frac{ k_{B}T}{8\pi\eta}
\ln\frac{t_{W}}{t_{\nu}} \\\nonumber
&=& D_E + \frac{ k_{B}T}{4\pi\eta}
\ln\frac{L}{\sigma_{cs}} \\
\end{eqnarray}\nonumber
which indicates that the diffusion of a particle in confinement should
scale with the log of the ratio of its radius to the pipe width.

We performed simulations to check the validity of this simple scaling argument.
The results are shown in Fig.~\ref{Fig:diffpipe222}, and can be accurately fitted to
Eq.(\ref{eq:pipe}).

While Eq.\ref{eq:pipe} works very well for the larger boxes, it
overestimates the diffusion coefficient for the smaller boxes.  This
is because the more complex wall effects shown in
Fig~\ref{Fig:normtails1} come into play so that the VACF no longer
shows simple $t^{-1}$ behavior assumed in Eq.\ref{eq:pipe}.
 We also note that in the
smallest systems studied, the Enskog contribution is more than half of the
overall diffusion (although this is not the reason for the deviation
from the simple $\ln[L/\sigma_{cs}]$ scaling).

It was shown by Bungay and Brenner~\cite{Bung73} using standard
methods of low Re number hydrodynamics, that the diffusion coefficient
of a sphere in a 3d pipe of radius $R_p$ drops rapidly with $R_p/R_c$.
Here we see from direct simulations of the VACF and the diffusion
coefficient that the same behavior can be seen in 2d.  It also drops as
a function of decreasing pipe width. It would be interesting to see if similar
hydrodynamic arguments to those used by Bungay and
Brenner~\cite{Bung73} could be used to explain the more rapid decrease
of the diffusion coefficient observed in 2d for stronger confinement.

\section{Conclusion}

We have applied the SRD simulation method to the study of the dynamics
of two dimensional disks in confined geometries. We calculated the
VACF for colloids and observed the predicted $t^{-1}$ behavior as
well as the more complex oscillating behavior and negative tails in
strong confinement. We also observed the logarithmic dependence of the
diffusion coefficient on system size, as originally predicted by
Saffman~\cite{Saff76} for the lateral diffusion of a cylinder in a
film.

Although the Saffman result describes the motion of a disk of
thickness $h$, and our simulation deals with disks, we can still map
our results onto a real physical system by equating the diffusion
coefficient measured in our simulations to that measured in
experiment.

Through this study we have shown that SRD can be fruitfully used to
simulate colloids in two dimensions.  This suggests that it could
easily be adapted for the study of other problems, such as protein and
lipid molecules in biological
membranes~\cite{Cone72,Cone73,Cone74,Edidin74,Pete82}, liquid domains in
giant unilamellar vesicles (GUVs)~\cite{Cicu07}, or colloids in a
liquid film~\cite{Cheu96,DiLe08}, or various examples from
microfluidics~\cite{Squi05}.

\acknowledgments
J.S. thanks Schlumberger Cambridge Research and IMPACT FARADAY for an
EPSRC CASE studentship which supported this work. A.A.L. thanks the
Royal Society (London) and 
J.T.P. thanks the Netherlands Organization for Scientific
Research (NWO) for financial support.  We thank E. Boek and I. Pagonabarraga for helpful conversations.

\end{document}